\documentstyle[12pt,epsfig]{article}

\newcommand{\dr}{\displaystyle \partial R}
\newcommand{\dt}{\displaystyle \partial T}
\newcommand{\dh}{\displaystyle \partial H}
\begin{document}

\large

\begin{center}
\Large \bf 
         Magnetically and thermally modulated \\
         microwave absorption \\
         in  $\rm\bf  Y_1 Ba_2 Cu_3 0_{7-x} $ single crystal near 
        $\rm\bf T_c $ \\
\end{center}

\vspace{2mm}

\begin{center}
\bf M.K.Aliev$^{a} $ , 
G.R.Alimov$^{a,} $\footnote[1]{Corresponding author. 
e-mail: gleb@iaph.silk.org }, I.Kholbaev$^{a} $,
L.I.Leonyuk$^{b} $,\\ T.M.Muminov$^{a} $,
B.A.Olimov$^{a} $, R.F.Rumi$^{c} $, H.I.Turkmenov$^{a} $ 
\end{center}

\begin{center}
\small $^{a} $Institute of Applied Physics, Tashkent State University
700095, Tashkent, Uzbekistan \\

$^{b} $Moscow State University, 117234, Moscow, Russia \\

$^{c} $Institute of Material Science, NPO "Physics-Sun",
    700084, Tashkent, Uzbekistan
\end{center}

\begin{abstract}
The microwave absorption R in the $ Y_1 Ba_2 Cu_3 O_{7-x} $ single crystals 
was investigated near $ T_c  \approx 92 K  $ and in the external magnetic 
field $ 0 < H \leq 9 kOe $ kOe. A modified ESR spectrometer was used in 
the experiment. The method of temperature modulation, along with the 
usual method of magnetic-field modulation, was  first applied in studying 
of the microwave response of these crystals. Peaks in the temperature 
dependencies of the signals $\dr /\dh $ and $\dr /\dt $ observed in the 
vicinity of $ T_c $ were differently shaped and slightly shifted one with
respect to another. The evolution of the peaks with variation of the 
magnetic field and angle $\theta $ between $\bf H $ and the $\bf c $-axis 
was traced. It has been shown that the observed difference of the 
temperature dependencies of the derivatives $\dr /\dh $ and $\dr /\dt $ 
occures due to the field-induced broadening of the superconducting 
transition, which is inherent in the high-$ T_c $ superconductors.
\end{abstract}

\vspace{1mm}

%%%%%%%%%%%%%%%%%%%%%%%%%%%%%%%%%%%%%%%%%%%%%%%%%%%%%%%%%%%%%%%%%%%%%%%%%%%
\section{Introduction}
\indent

It is well known that the superconducting transition 
in the high-$ T_c $ superconductors shows a dramatic 
\cite{ref1} field-induced broadening. In particular this phenomenon 
prevents from determining the upper critical field 
$ H_{c_2} (T) $ for high $ T_c $ superconductors  by traditional methods.
The origin of the broadening is still unclear and remains a subject of 
investigations \cite{ref2} - \cite{ref10}.

As a source of information on superconducting transition
properties the results of microwave measurements performed 
by technique of ESR spectrometers may be useful. 
It has been found \cite{ref11} - \cite{ref15} 
that temperature dependence of the 
microwave response of high-$ T_c $ superconductors, recorded
by an ESR spectrometer with modulating magnetic 
field, exhibits a peak near $ T_c $. It should be noted that
in this case the measured signal represents the derivative  
$\dr /\dh $, where R is a microwave absorption. In this connection
it has been supposed \cite{ref12,ref14} that the
temperature dependence of $\dr /\dh $ in the vicinity of
$ T_c $ is similar to that of $\dr /\dt $, 
as it occurs for traditional
superconductors, and so the observed $\dr /\dh $ -peak may be used for
analysis of superconducting transition in the 
high- $ T_c $ superconductors.

An original point of the present paper is that we have fulfilled
the direct measurement of the derivative $\dr /\dt $
for a single crystal Y-Ba-Cu-O and established 
the essential difference of its temperature dependence from that
of the derivative $\dr /\dh $.
It is shown that this difference is due to the field-induced
broadening of superconducting transition.

%%%%%%%%%%%%%%%%%%%%%%%%%%%%%%%%%%%%%%%%%%%%%%%%%%%%%%%%%%%%%%%%%%%%%%%%%%%
\section{Experimental}
\indent
 
In the experiment the ESR spectrometer SE/X-2543 "Radiopan"
with the X-band $ TE_{102} $ resonant cavity 
($\nu = 9.5 GHz, Q = 5000) $ 
was employed.

A usual field arrangement with 
$ {\bf H}\bot {\bf H_1} $, 
where H is the static field
and  $ {\bf H_1}$ - the microwave  field 
was used  (see insert in fig. 2).  
The  crystals  were  oriented  with 
$ {\bf c}\bot {\bf H_1}$. The angle $\theta $ between the 
$ {\bf c} $ and ${\bf H} $ could 
be varied by rotating the crystal
around the ${\bf H_1}$ direction and could be 
fixed with an accuracy of $ \pm 1^\circ $. 

In order to investigate the superconducting transition at low fields
some modifications are made because the electromagnet of ESR spectrometer
has remnant magnetic field with $\sim 30 Oe $. The resonator is taken out
from the large electromagnet and for creating low fields a pair 
of Helmholtz coils is used. To ensure that the field is zero 
at the zero current in the coils, the resonator and coils were placed
inside the cylindrical permalloy screen, reducing the geomagnetic
field more than hundred times.

A light-beam assisted temperature-control system operating within a 
temperature range from 77 to 180 K was specially designed for this
experiment \cite{ref16}. The advantage of this system is the feasibility 
of modulating the temperature of a sample with a frequency of 
80 Hz and amplitude of $ 10^{-2} - 10^{-1} $ K. The rms temperature
instability over a 5-min time interval is within 0.06 K, the 
temperature gradient in a sample is - 0.01 K/mm for $ T\sim 90 $ K,
and system relaxation time is 1-10 s.

The lock-in detection was used for recording the modulated microwave 
absorption for the following cases: 
\begin{itemize} 
\item[a)] at frequency 100 kHz when
modulating magnetic field was applied
(maximum modulation amplitude $ h_{max} $ =10 Oe ),
\end{itemize}
\begin{itemize} 
\item[b)] at frequency 80 Hz 
in the case of modulating the
temperature of a sample.
\end{itemize}
It should be noted 
that a direct measurement of the 
temperature modulation amplitude represents a rather 
complicated problem. Here we 
can point that, according to ref. \cite{ref16}, its value 
can be chosen in the range $ 10^{-2} - 10^{-1} K$, 
and that its variation over the transition region does 
not exceed 10\%. Since the 
amplitude is much less than the transition width, 
the observed signal can be considered 
as the derivative $\dr /\dt $. 

All the measurements were carried out at the constant 
microwave power level of -17 dB (130 mW). 

The single crystals were grown from the melt 
$ Y_{0,99} Ba_{2,00} Cu_{2,89} O_{7-x} $. The 
mixture was heated to $ 1100^\circ C $ in the $ ZrO_2 :Y $ crucible 
at the rate $ 2-5 ^\circ C/h $, and then 
cooled to the room temperature at the same rate. 
The crystal composition was 
determined by the method of X-ray micro-spectroscopy 
and corresponded to the 
formula $ Y_{0,99} Ba_{2,00} Cu_{2,89} O_{7-x} $. 
The lattice parameters, as determined by the X-ray 
diffraction, were: 
$ a=3,85A^{\phantom{H}^{\phantom{H}^
{\!\!\!\!\!\!\!\!\!\!\!\!\!\!\!\!\! \circ }}} $,
$ b=3,89A^{\phantom{H}^{\phantom{H}^
{\!\!\!\!\!\!\!\!\!\!\!\!\!\!\!\!\! \circ }}} $ and
$ c=11,74A^{\phantom{H}^{\phantom{H}^
{\!\!\!\!\!\!\!\!\!\!\!\!\!\!\!\!\! \circ }}} $.
The crystals were shaped as 
thin plates with thickness $ d\sim 0,03 mm $ 
along the $ {\bf c} $ 
and area of $ <1 mm^2 $ in the $({\bf a}, {\bf b}) $
plane.

%%%%%%%%%%%%%%%%%%%%%%%%%%%%%%%%%%%%%%%%%%%%%%%%%%%%%%%%%%%%%%%%%%%%%%%%%%%
\section{Experimental results and discussion}
\indent

The crystal samples exhibiting simple-shaped 
single peak in the temperature 
dependence of the derivatives $\dr /\dh $ and $\dr /\dt $ 
were selected for detailed 
studies. The so called "low-field" signals \cite{ref17,ref18}  
were absent in all samples. These 
conditions ensured that the observed temperature 
dependence of the measured signal was 
determined exclusively by the superconducting transition. 
It should be stressed also, 
that no irreversible effects were observed in the studied samples.

In general 5 samples were investigated. 
All the results obtained in the experiment 
might be summarized in 
several regularities demonstrated in figs. 1-3 
(all figures refer to the same sample). The 
most important result  is the qualitative 
difference of the temperature 
dependencies of the derivatives 
$\dr /\dh $ and $\dr /\dt $. It is seen 
in fig. 1, that the 
corresponding peaks are shifted one with respect to another  
in the temperature scale.
%%%%%%%%%%%%%%%%%%%%%%%%%%%%%%%%%%%%%%%%%%%%%%%%%%%%%%%%%%%%%%%%%%
\begin{figure}
\vskip4cm
\begin{center}
\hspace*{.5cm}
\parbox{8cm}{\epsfxsize=6.cm \epsfysize=8.cm \epsfbox [5 5 500 500]
{fig1.eps}{}}
\vskip -3cm
\end{center}
\begin{center}
\parbox{8.5cm}{\small 
Fig.1. Temperature dependencies of the derivatives 
$\dr /\dh $ and $\dr /\dt $ at $\theta = 0 $ and different H.}
\end{center}
\end{figure}
%%%%%%%%%%%%%%%%%%%%%%%%%%%%%%%%%%%%%%%%%%%%%%%%%%%%%%%%%%%%%%%%%%
Moreover, the $\dr /\dh $  peak
is broader than the $\dr /\dt $ peak and has
an asymmetric shape with a clearly seen 
low temperature tail. 
A different behavior of these peaks with
increasing external field is also seen in fig. 1.
The amplitude of the $\dr /\dh $ peak 
decreases faster than that 
of the $\dr /\dt $ one, and shows a larger shift toward lower 
temperature. 

We have also 
studied an asymptotic behavior with the decreasing external field, 
$ H\to 0$. It was found 
that below 20 Oe the amplitude and position of the 
$\dr /\dh $ peak remain constant, in the range of
error limits, and its shape became more symmetrical with decreasing 
field, approaching 
that of the $\dr /\dt $ peak. The latter did not experience 
any changes down to the lowest 
attainable field strength of $ 10^{-3}\div 10^{-2}$ Oe. 
  
Measurements of the derivative $\dr /\dh $
at $ H\to 0 $ are subject to a specific 
problem. The corresponding
$\Delta R $ signal was recorded at the values of field 
modulation amplitude h compatible with 
the condition $\Delta R \sim (\dr /\dh)\cdot h $. 
Special tests showed that this condition 
fulfilled as long as $ h<H $. The latter 
condition made it difficult to carry out measurements 
at $ H<2,5Oe $, and at H=1 Oe  
the $\Delta R $  signal fell to the noise level. This should be kept in 
mind when considering our 
conclusion about the constancy of the $\dr /\dh $ peak 
amplitude and position at $ H\to 0 $. 
  
The  $\dr /\dh $ and $\dr /\dt $ peaks differently 
responded to the variation of the angle  
$\theta $ from $ 0^\circ $ to $ 90^\circ $. 

At low fields, $ H\leq 20 Oe $, the $\dr /\dt $
peak was practically constant in the range 
$ 0^\circ \leq\theta\leq 90^\circ $, whereas 
a significant changes of the  $\dr /\dh $ 
peak features were observed. The
amplitude of $\dr /\dh $ peak decreased 
gradually, reaching at $\theta=90^\circ $ about 1/10 
of its initial value, while its position   
shifted slightly toward higher temperature (see fig. 2).
At high fields these peaks showed 
a reverse behavior: the $\dr /\dt $  peak 
changed significantly (see fig. 3), 
whereas the $\dr /\dh $ peak remains unchanged.

%%%%%%%%%%%%%%%%%%%%%%%%%%%%%%%%%%%%%%%%%%%%%%%%%%%%%%%%%%%%%%%%%%
\begin{figure}
\vskip2cm
\begin{center}
\hspace*{-4.cm}
\parbox{6cm}{\epsfxsize=6.cm \epsfysize=5.6cm \epsfbox [5 5 500 500]
{fig2.eps}{}}
\hspace*{2.cm}
\parbox{6cm}{\epsfxsize=6.cm \epsfysize=5.6cm \epsfbox[5 5 500 500]
{fig3.eps}{}}
\vskip -0.5cm
\end{center}
\begin{center}
\parbox{7.cm}{\small 
Fig.2. Evolution of the temperature dependence of the derivative 
$\dr /\dh $ at H = 20 Oe 
with changing angle $\theta $.
The geometrical arrangement of 
the experiment is shown in the insert.}
\hspace*{1.cm}
\parbox{7.cm}{\small 
\vskip -0.8cm
Fig.3. Evolution of the temperature dependence of the 
derivative $\dr /\dt $ 
at H = 9 kOe with changing angle $\theta$.}
\end{center}
\end{figure}
%%%%%%%%%%%%%%%%%%%%%%%%%%%%%%%%%%%%%%%%%%%%%%%%%%%%%%%%%%%%%%%%%%

Common features of these
two peaks were their narrowing and shifting to 
higher temperature (for about 1 K and 1.5 K 
correspondingly at H =9 kOe)
when increasing $\theta $ to $ 90^\circ $. 
  
We preface the discussion 
of the results with the following remark. 
The absorption function R(H,T) satisfies the 
identity:

\begin{equation}  \label{f1}
  \dr / \dt =-(dH_r(T)/dT)\cdot (\dr / \dh),
\end{equation}
where $ H_r (T)$ is determined by the equation 

$$  
  R (H_r(T),T)=r=const.
$$
The function $ H_r (T)$ describes the "equiabsorption" 
line in the 
phase plane (H, T). Such a line in the 
region of the superconducting transition is 
commonly interpreted as the temperature dependence  
of the upper critical field $ H_{c_2} (T)$. This 
interpretation is valid provided that the 
dependence of the line's slope $  dH_r / dT $ on the absorption 
level r is weak in the temperature interval of the transition
at any values of H
(otherwise, the lines
$  H_r(T) $ with different r would diverge in the phase plane, 
and choosing of $  H_r (T) $ for $ H_{c_2} (T)$ would be 
uncertain). Usually, this condition is fulfilled 
in the low-temperature superconductors
and in this case, according to eq.(1), the temperature 
dependencies of $\dr /\dh $ and $\dr /\dt $ have to be 
similar. Note, that this is possible only if 
the curve $ R(H=const,T)$,describing the superconducting
transition, is shifted as a whole with
changing field H.

It is seen in fig.1 that 
the peaks in the temperature dependencies 
of the derivatives $\dr /\dh $ and $\dr /\dt $, 
for a given high-$ T_c $ superconductor,
are differently shaped, 
and, which is more important, are shifted 
one with respect to another in the 
temperature scale. This means, according to eq. (1), 
that the value of $ dH_r / dT $ is not uniform 
in the transition region and therefore should be 
essentially dependent on the absorption 
level. Hence, the relative shift of the peaks seen in 
the graphs of fig.1 is an evidence of the field-induced
broadening  of the superconducting transition. It should be 
underlined also, that the comparative analysis of 
the peaks  $\dr /\dh $ and $\dr /\dt $ allows one 
to observe the transition's broadening even at very low fields. 
A direct observation of 
this effect would require a much higher precision 
of temperature measurements. Thus, 
we hope that the results described above may be used 
in testing the existing models of 
the field-induced broadening in high-$ T_c $ superconductors. 

\vspace{4mm}
\noindent {\Large\bf Acknowledgements} 
\vspace{2mm}

Authors are greatly indebted to B.Yu.Sokolov 
for valuable discussions, and to 
I.R.Mikulin for the assistance in the design 
of the temperature monitoring system.


\begin{thebibliography}{15}
%%%%%%%%%%%%%%%%%%%%%%%%%%%%%%%%%%%%%%%%%%%%%%%%%%%%%%%%%%%%%%%%%%%%%%%%%%%
\bibitem{ref1} Y. Iye, T. Tamegai, H. Takeya and Takei, in:
Superconducting Materials, eds. S. Nakajima and H. Fakuyama, 
Jpn. J. Appl. Phys. Series 1 (Publication Office, Japanese 
Journal of Applied Physics, Tokyo, 1988)p.46.
%%%%%%%%%%%%%%%%%%%%%%%%%%%%%%%%%%%%%%%%%%%%%%%%%%%%%%%%%%%%%%%%%%%%%%%%%%%
\bibitem{ref2} Y. Yeshurun , A.P. Malozemoff, Phys. Rev. Lett. 
60 (1988) 2202.
%%%%%%%%%%%%%%%%%%%%%%%%%%%%%%%%%%%%%%%%%%%%%%%%%%%%%%%%%%%%%%%%%%%%%%%%%%%
\bibitem{ref3} M. Tinkham,  Phys. Rev. Lett. 61 (1988) 1658.
%%%%%%%%%%%%%%%%%%%%%%%%%%%%%%%%%%%%%%%%%%%%%%%%%%%%%%%%%%%%%%%%%%%%%%%%%%%
\bibitem{ref4} T.T.M. Palstra, B. Batlogg, R.B. van Dover, L.F. Schneemeyer,
J.V. Waszczak, Phys. Rev. B41 (1990) 6621.
%%%%%%%%%%%%%%%%%%%%%%%%%%%%%%%%%%%%%%%%%%%%%%%%%%%%%%%%%%%%%%%%%%%%%%%%%%%
\bibitem{ref5} K.H. Lee, D. Stroud,  Phys. Rev.  B46 (1992) 5699.
%%%%%%%%%%%%%%%%%%%%%%%%%%%%%%%%%%%%%%%%%%%%%%%%%%%%%%%%%%%%%%%%%%%%%%%%%%%
\bibitem{ref6} C.C. Chin and T. Morishita, Physica C 207 (1993) 37. 
%%%%%%%%%%%%%%%%%%%%%%%%%%%%%%%%%%%%%%%%%%%%%%%%%%%%%%%%%%%%%%%%%%%%%%%%%%%
\bibitem{ref7} H.A. Blackstead, Phys. Rev.  B47 (1993) 11411. 
%%%%%%%%%%%%%%%%%%%%%%%%%%%%%%%%%%%%%%%%%%%%%%%%%%%%%%%%%%%%%%%%%%%%%%%%%%%
\bibitem{ref8} H.A. Blackstead, G.A. Kapustin, Physica C219 (1994) 109.  
%%%%%%%%%%%%%%%%%%%%%%%%%%%%%%%%%%%%%%%%%%%%%%%%%%%%%%%%%%%%%%%%%%%%%%%%%%%
\bibitem{ref9} H.A. Blackstead, D.B. Pulling, M. Paranthaman, 
J. Brynestad, Phys. Rev. B51 (1995) 3783.
%%%%%%%%%%%%%%%%%%%%%%%%%%%%%%%%%%%%%%%%%%%%%%%%%%%%%%%%%%%%%%%%%%%%%%%%%%%
\bibitem{ref10} X.-J. Xu, L. Fu and Y.-H. Zhang, Physica C 282-287 
(1997) 1557.
%%%%%%%%%%%%%%%%%%%%%%%%%%%%%%%%%%%%%%%%%%%%%%%%%%%%%%%%%%%%%%%%%%%%%%%%%%%
\bibitem{ref11} K. Moorjani, J. Bohandy, F.J. Adrian, 
B.F. Kim, R.O. Shull, C.K. Chiang, 
L.J. Swartzendruber, L.H. Bennett, Phys. Rev. B36
(1987) 4036.
%%%%%%%%%%%%%%%%%%%%%%%%%%%%%%%%%%%%%%%%%%%%%%%%%%%%%%%%%%%%%%%%%%%%%%%%%%%
\bibitem{ref12} B.F. Kim, J. Bohandy, K. Moorjani, F.J.  Adrian, 
J. Appl. Phys.  63 (1988) 2029.                                       
%%%%%%%%%%%%%%%%%%%%%%%%%%%%%%%%%%%%%%%%%%%%%%%%%%%%%%%%%%%%%%%%%%%%%%%%%%%
\bibitem{ref13} M.K. Aliev, J. Wawryshchuk, S.P. Wolosyaniy,T.M. Muminov,
B. Olimov and I. Kholbaev, Fiz. Tverd. Tela (Leningrad) 31 (1989) 254.
%%%%%%%%%%%%%%%%%%%%%%%%%%%%%%%%%%%%%%%%%%%%%%%%%%%%%%%%%%%%%%%%%%%%%%%%%%%
\bibitem{ref14} D. Shaltiel, H. Bill, A. Grayevsky, A. Junod, D. Lovy,
W. Sadowski, E. Walker,  Phys. Rev. B 43  (1991) 13594.
%%%%%%%%%%%%%%%%%%%%%%%%%%%%%%%%%%%%%%%%%%%%%%%%%%%%%%%%%%%%%%%%%%%%%%%%%%%
\bibitem{ref15} S.G. L'vov, Yu.I. Talanov, R.I. Khasanov and V.A. Shustov,
(russian) Sverkhprovodimost: Fiz. Khim. Tekh. 6  (1993) 1175.
%%%%%%%%%%%%%%%%%%%%%%%%%%%%%%%%%%%%%%%%%%%%%%%%%%%%%%%%%%%%%%%%%%%%%%%%%%%
\bibitem{ref16} M.K. Aliev, G.R. Alimov, T.M. Muminov, B. Olimov,
B.Yu. Sokolov, R.R. Usmanov and I.Kholbaev, Instruments 
and Experimental Techniques, 39  (1996) 769 (Translated from Pribory
i Tekhnika Eksperimenta 5 (1996) 152), (E-preprint, cond-mat/9809372).
%%%%%%%%%%%%%%%%%%%%%%%%%%%%%%%%%%%%%%%%%%%%%%%%%%%%%%%%%%%%%%%%%%%%%%%%%%%
\bibitem{ref17} K.W. Blazey, A.M. Portis, K.A. Muller, F.H. Holtzberg,
Europhys. Lett. 6 (1988) 457.
%%%%%%%%%%%%%%%%%%%%%%%%%%%%%%%%%%%%%%%%%%%%%%%%%%%%%%%%%%%%%%%%%%%%%%%%%%%
\bibitem{ref18} A. Dulcic, R.H. Crepeau, 
J.H. Freed, Phys. Rev. B38 (1988) 5002.
\end{thebibliography}
\end{document}